\documentclass[10pt, twocolumn]{article}

\usepackage[TS1,OT1]{fontenc}
\usepackage{xspace, graphicx,units, cite, mcite, fix2col}
\usepackage[nolists,nomarkers]{endfloat}

\newcommand*{\figref}[1]{Fig.~\ref{fig:fig#1}\xspace}
\newcommand*{\figpart}[1]{{\sffamily (#1)}}
\newcommand*{\bfield}{\ensuremath{B}\xspace}
\newcommand*{\Vdot}{\ensuremath{V_{\mathrm{dot}}}\xspace}
\newcommand*{\Ibias}{\ensuremath{I_{\mathrm{bias}}}\xspace}
\newcommand*{\Vga}{\ensuremath{V_{g1}}\xspace}
\newcommand*{\Vgb}{\ensuremath{V_{g2}}\xspace}

\newcommand*{\ph}{\ensuremath{\phi}\xspace}
\newcommand*{\Ao}{\ensuremath{A_{0}}\xspace}
\newcommand*{\Bo}{\ensuremath{B_{0}}\xspace}
\newcommand*{\temp}{\ensuremath{T}\xspace}
\newcommand*{\Aac}{\ensuremath{A_{\mathrm{ac}}}\xspace}
\newcommand*{\Aacsq}{\ensuremath{A_{\mathrm{ac}}^{2}}\xspace}
\newcommand*{\Aacro}{\ensuremath{A_{\mathrm{ac}}^{1/2}}\xspace}
\newcommand*{\Xa}{\ensuremath{X_{1}}\xspace}
\newcommand*{\Xb}{\ensuremath{X_{2}}\xspace}
\newcommand{\comment}[1]{\relax}
\newcommand*{\avg}[1]{\ensuremath{\left<#1\right>}\xspace}
\newcommand*{\adot}{\ensuremath{a_{\mathrm{dot}}}\xspace}

\newcommand*{\emmisiv}{\ensuremath{dn/dX}\xspace}
\newcommand*{\ar}{\ensuremath{\alpha}\xspace}
\newcommand*{\sAo}{\ensuremath{\sdev(\Ao)}\xspace}
\newcommand*{\usim}{\ensuremath{\mathbin\sim}}

\newcommand*{\micro}{\ensuremath{\mu}}

\makeatletter
\def\sdev{\mathop{\operator@font \sigma}\nolimits}
\makeatother

\newcommand{\figonecaption}{%
	\figpart{A} Pumped dc voltage \Vdot as a function of the phase 
	difference \ph between two shape-distorting ac voltages and 
	magnetic field \bfield.  Note the sinusoidal dependence on \ph and 
	the symmetry about $\bfield = 0$ (dashed white line).  \figpart{B} 
	$\Vdot (\phi)$ for several different magnetic fields (solid 
	symbols) along with fits of the form $\Vdot = \Ao \sin \phi + \Bo$ 
	(dashed curves).  \figpart{C} Schematic of the measurement setup 
	and micrograph of device 1. \Ibias is set to 0 for pumping 
	measurements.}

\newcommand{\figtwocaption}{%
	\figpart{A} Standard deviation of the pumping amplitude, \sAo, as 
	a function of ac pumping frequency.  The slope is $\sim 
	\unit[40]{nV/MHz}$ for both device 2 (solid symbols) and 3 (open 
	symbols).  Circular symbols represent a second set of data taken 
	for device 3.  \figpart{B} A typical data set corresponding to one 
	point in \figpart{A}, along with fit parameters \Ao (open bars) 
	and \Bo (solid bars) for each configuration.}

\newcommand{\figthreecaption}{%
	Standard deviation of the pumping amplitude, \sAo, as a function 
	of the ac driving amplitude \Aac, along with fits to $\sAo \propto 
	\Aacsq$ below \unit[80]{mV} (dashed line), $\sAo \propto \Aac$ 
	(solid line), and $\sAo \propto \Aacro$ (dotted line) above 
	\unit[80]{mV}.  Lower Inset: Sinusoidal dependence of 
	$\Vdot(\phi)$ at small and intermediate \Aac (solid curve, $\Aac = 
	\unit[100]{mV}$) becomes nonsinusoidal for strong pumping (dotted 
	curve, $\Aac = \unit[260]{mV}$), but maintains $\Vdot(\pi) = 0$, 
	required by time-reversal symmetry.  Upper Inset: Schematic of the 
	loop swept out by the pumping parameters \Xa and \Xb.  The charged 
	pumped per cycle can be written in terms of an integral over the 
	surface $\alpha$ enclosed by the loop.}

\newcommand{\figfourcaption}{%
	Pumping amplitude \sAo as a function of temperature \temp with a 
	power law fit (dashed line).  At lower temperatures, there 
	is a rounding off of the \temp dependence, consistent with an 
	expected saturation when lifetime broadening exceeds temperature 
	below $\usim \unit[100]{mK}$.}

\newcommand{\figfivecaption}{%
	\figpart{A} Gray-scale plot of $\Vdot(\phi = \pi/2)$ as a function 
	of magnetic field \bfield and dc gate voltage \Vga 
	\protect\cite{note:bfieldshift}.  Note the symmetry and the 
	characteristic scales of pumping fluctuations.  \figpart{B} 
	Average (dotted curve) and standard deviation (solid curve) of 36 
	uncorrelated samples of $\Vdot(\phi = \pi/2)$ as a function of 
	\bfield measured at different dc gate voltages, \Vga and \Vgb [not 
	the same data set as (A)].  The average is small and fluctuates 
	around zero.  The standard deviation shows a peak around $\bfield 
	= 0$ roughly twice its value away from zero with a width 
	corresponding to roughly one quantum of flux, $h/e$, through the 
	dot.  The reduction of pumping fluctuations away from $\bfield = 
	0$ is presumably associated with the breaking of time-reversal 
	symmetry, similar to the reduction of conductance fluctuations 
	seen upon breaking time-reversal symmetry 
	\protect\cite{baranger:mesoscopic, *jalabert:universal, 
	*chan:ballistic}.}

\title{An Adiabatic Quantum Electron Pump}
\date{}

\author{M. Switkes\textsuperscript{*}, C. M. Marcus\textsuperscript{*},
             K. Campman\textsuperscript{\textdagger},
	  A. C. Gossard\textsuperscript{\textdagger}\\
	  \textsuperscript{*}Department of Physics, Stanford University, 
	  Stanford CA 94305\\
	  \textsuperscript{\textdagger}Materials Department, University of California,
	  Santa Barbara, CA 93106}

\begin{document}

\twocolumn[%
\maketitle%
\begin{center}%
	\parbox{0.9\textwidth}{%
	A quantum pumping mechanism which produces dc current or voltage 
	in response to a cyclic deformation of the confining potential in 
	an open quantum dot is reported.  The voltage produced at zero 
	current bias is sinusoidal in the phase difference between the two 
	ac voltages deforming the potential and shows random fluctuations 
	in amplitude and direction with small changes in external 
	parameters such as magnetic field.  The amplitude of the pumping 
	response increases linearly with the frequency of the deformation.  
	Dependencies of pumping on the strength of the deformations, 
	temperature, and breaking of time-reversal symmetry are also 
	investigated.}%
	\vspace*{\baselineskip}
	\rule{0.25\textwidth}{1.5pt}
\end{center}
]

\noindent
Over the past decade, research into the electrical transport 
properties of mesoscopic systems has provided insight into the quantum 
mechanics of interacting electrons, the link between quantum mechanics 
and classical chaos, and the decoherence responsible for the 
transition from quantum to classical physics \cite{beenakker:quantum, 
kouwenhoven:electron}.  The majority of this research has focused on 
transport driven directly by an externally applied bias.  We present 
measurements of an adiabatic quantum electron pump, exploring a class 
of transport in which the flow of electrons is driven by cyclic 
changes in the wave function of a mesoscopic system.

A deformation of the confining potential of a mesoscopic system that 
is slow compared to the relevant energy relaxation times changes the 
wave function of the system while maintaining an equilibrium 
distribution of electron energies.  In systems connected to bulk 
electron reservoirs by open leads supporting one or more transverse 
quantum modes, the wave function extends into the leads and these 
adiabatic changes can transport charge to or from the reservoirs.  A 
periodic deformation that depends on a single parameter cannot result 
in net transport; any charge that flows during the first half period 
will flow back during the second.  On the other hand, deformations 
that depend on two or more parameters changing in a cyclic fashion can 
break this symmetry and in general can provide net transport.  This 
transport mechanism was originally described by Thouless 
\cite{thouless:quantization} for isolated (or otherwise gapped) 
systems at zero temperature.  The theory has recently been extended to 
open systems at finite temperature \cite{spivak:mesoscopic, 
zhou:mesoscopic, brouwer:scattering}.  Here, we present the first 
experimental investigation of this phenomenon.

Before characterizing the adiabatic quantum pump in the present 
experiment, it is useful to recall other mechanisms that produce a dc 
response to an ac driving signal in coherent electronic systems.  One 
mechanism relies on absorption of radiation to create a 
non-equilibrium distribution of electron energies, leading to 
photon-assisted tunneling \cite{kouwenhoven:photon, 
*kouwenhoven:observation} in systems with asymmetric tunneling leads, and a 
mesoscopic photovoltaic effect \cite{fal:mesoscopic, *liu:mesoscopic} 
in open systems.  A second mechanism, the classical analog of the 
quantum pumping measured in this experiment, has been observed in 
single \cite{kouwenhoven:quantized, *kouwenhoven:quantizedz} and 
multiple \cite{pothier:single} quantum dots in which transport is 
dominated by the Coulomb blockade \cite{kouwenhoven:electron}.  In 
this regime, the capacitive energy needed to add a single electron to 
the system is greater than the temperature and applied bias, 
blockading transport through the dot.  Electrons can be added one by 
one by changing the potential of the isolated dot relative to 
the reservoirs.  Each cycle begins by isolating the system from one 
electron reservoir---for example by increasing the height of one 
tunneling barrier---while forcing one or more electrons to enter from 
the other reservoir by changing the  potential in the system.  
The cycle continues, reversing the configuration to isolate the system 
from the reservoir that supplied the electrons, and forcing the extra 
electrons out into the other reservoir, yielding a net flow quantized 
in units of the electron charge times the frequency applied.  This 
cycle requires two ac control voltages with a phase difference between 
them.  The magnitude and direction of the pumping are determined by 
these voltages; there are no random fluctuations due to quantum 
effects.  The control and quantization of current provided by the 
Coulomb blockade pump has motivated its development for use as a 
precision current standard (see for example \cite{keller:rare}).

Adiabatic quantum pumping in open structures also requires two ac 
voltages, and produces a response that is linear in the ac frequency.  
However, because the system is open to the reservoirs, Coulomb 
blockade is absent and the pumping response is not quantized.  Quantum 
pumping is driven not by cyclic changes to barriers and potentials, 
but by shape changes in the confining potential or other parameters 
that affect the interference pattern of the coherent electrons in the 
device.

\begin{figure*}
   \begin{center}
   \includegraphics{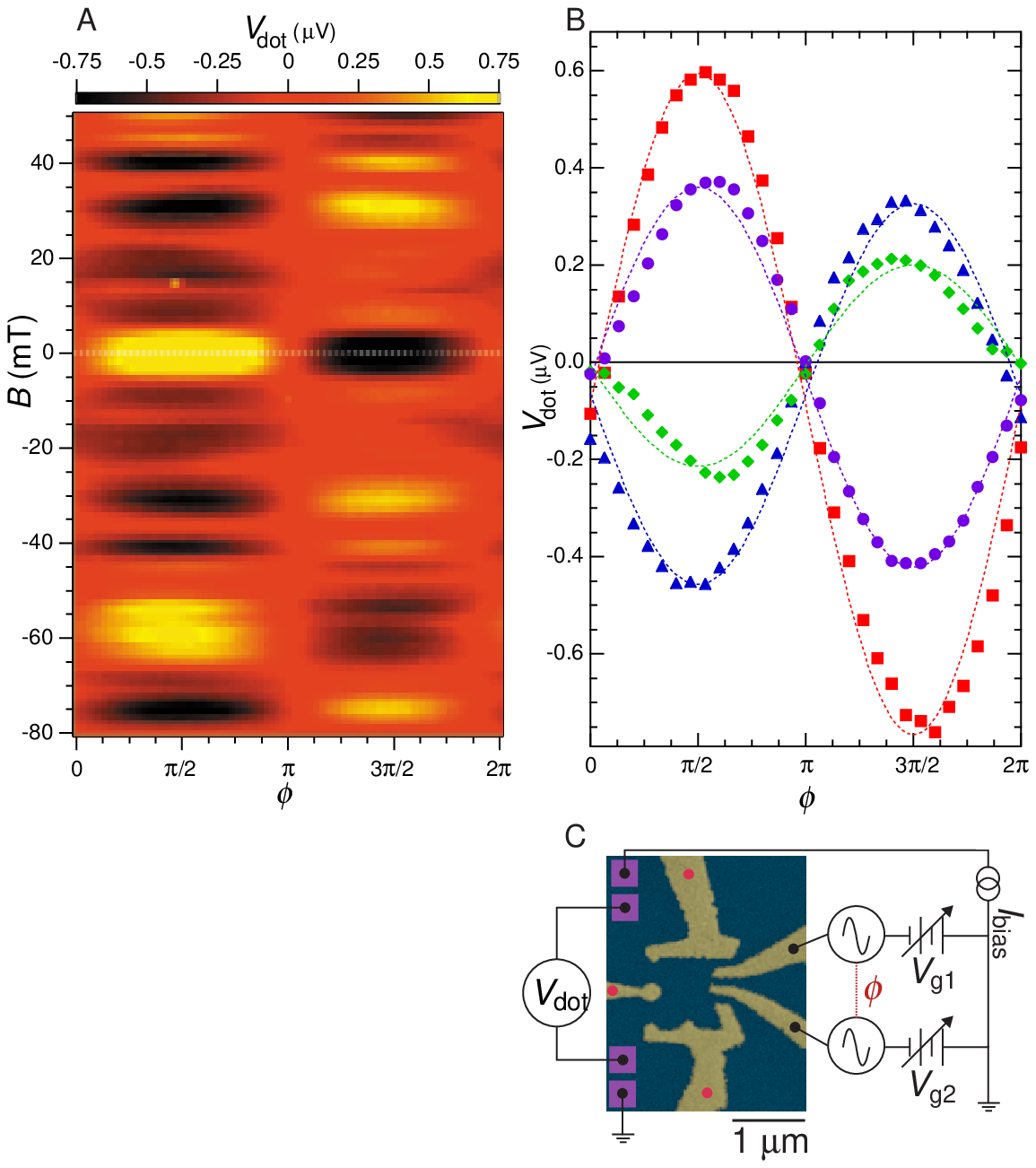}
   \caption{\figonecaption}%
   \label{fig:fig1}
   \end{center}
\end{figure*}

Many aspects of adiabatic quantum pumping can be understood in terms 
of emissivity, \emmisiv, which characterizes the number of electrons 
$n$ entering or leaving the device in response to a small change in 
some parameter $\delta X$, such as a distortion of the confining 
potential \cite{buttiker:current}.  The change in the charge of the 
dot is thus $\delta Q = e \sum \delta X_{i} \, \emmisiv_{i}$.  
Integrating the total emissivity along the closed path in the space of 
parameters $X_{i}$ defined by the pumping cycle then yields the total 
charge pumped during each cycle \cite{brouwer:scattering}.  For the 
particular case of pumping with two parameters (for example, shape 
distortions at two locations on the dot), the line integral can be 
written as an integral over the surface enclosed by the path, $Q 
\propto \int_{\ar} \xi \, d\Xa d\Xb$ \cite{brouwer:scattering} where 
$\xi$ depends on the emissivities at points in parameter space 
enclosed by the path.  Because changes in external parameters 
rearrange the electron interference pattern in the device, 
emissivities fluctuate randomly as parameters are changed, similar to 
the well-known mesoscopic fluctuations of conductance in coherent 
samples.  When the pumping parameters make an excursion smaller than 
the correlation length of the fluctuations of emissivity, $\xi$ 
remains essentially constant throughout the pumping cycle and the 
total charge pumped per cycle depends only on the area enclosed by the 
path in parameter space, \ar.  These straightforward observations 
explain many of the qualitative features of our data.

Measurements of adiabatic quantum pumping in three similar 
semiconductor quantum dots defined by electrostatic gates patterned on 
the surface of a GaAs/AlGaAs hetero\-structure using standard 
electron-beam lithography techniques are reported.  Negative voltages 
($\sim \unit[-1]{V}$) applied to the gates form the dot by depleting 
the two dimensional electron gas at the heterointerface \unit[56]{nm} 
(device~1) or \unit[80]{nm} (devices~2 and 3) below the surface.  All 
three dots have lithographic areas $\adot \sim \unit[0.5]{\micro 
m^{2}}$ giving an average single particle level spacing $\Delta= 2 \pi 
\hbar^{2}/m^{*}\adot \sim \unit[13]{\mu V} (\approx \unit[150]{mK})$.  
The three devices show similar behavior, and most of the data 
presented here are for device~3.  In the micrograph of device~1 
(\figref{1}{C}), the three gates marked with red circles control the 
conductances of the point-contact leads connecting the dot to 
electronic reservoirs.  Voltages on these gates are adjusted so that 
each lead transmits $N \sim 2$ transverse modes, giving an average 
conductance through the dot $g \sim 2e^{2}/h$.  The remaining two 
gates are used to create both periodic shape distortions necessary for 
pumping and static shape distortions that allow ensemble averaging 
\cite{baranger:mesoscopic, *jalabert:universal, *chan:ballistic, 
note:pumping}.

Except where noted, measurements were made at a pumping frequency $f = 
\unit[10]{MHz}$, base temperature $\temp = \unit[330]{mK}$, dot 
conductance $g \sim 2e^{2}/h \approx (\unit[13]{k\Omega})^{-1}$, and 
ac gate voltage $\Aac = \unit[80]{mV}$ peak-to-peak.  For comparison, 
the gate voltage necessary to change the electron number in the dot by 
one is $\sim \unit[5]{mV}$.  Measurements were carried out over a 
range of magnetic field from 30 to 80 mT, which allows several quanta 
of magnetic flux, $\varphi_{0}=h/e$, to penetrate the dot 
($\varphi_{0} / \adot \sim \unit[10]{mT}$) while keeping the classical 
cyclotron radius much larger than the dot size 
($r_{\mathrm{cyc}} \unit{[\micro m]} \sim 80/\bfield 
\unit{[mT]}$).

The general character of quantum pumping, including antisymmetry about 
phase difference $\phi = \pi$, sinusoidal dependence on $\phi$ (for 
small amplitude pumping), and random fluctuations of amplitude as a 
function of perpendicular magnetic field is illustrated in \figref{1}.  
The pumping amplitude is quantified by the values $\Ao$ and $\Bo$ 
which are extracted from fits of the form $\Vdot(\phi) = \Ao \sin\phi 
+ \Bo$ (shown as dotted lines in \figref{1}{B}).

\begin{figure}
    \begin{center}
    \includegraphics{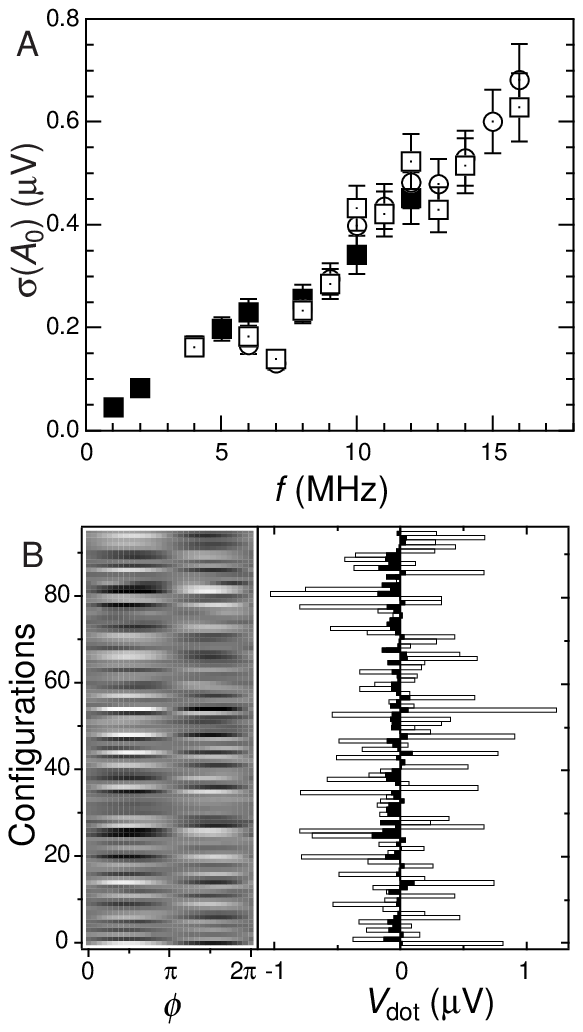}
	\caption{\figtwocaption}%
    \label{fig:fig2}
    \end{center}
\end{figure}

\begin{figure}
    \begin{center}
	\includegraphics{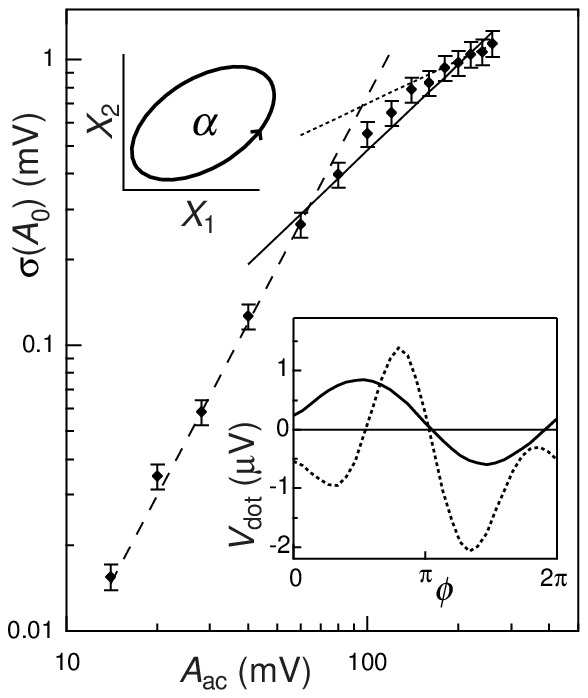}
	\caption{\figthreecaption}%
	\label{fig:fig3}
    \end{center}
\end{figure}

\begin{figure}
    \begin{center}
	\includegraphics{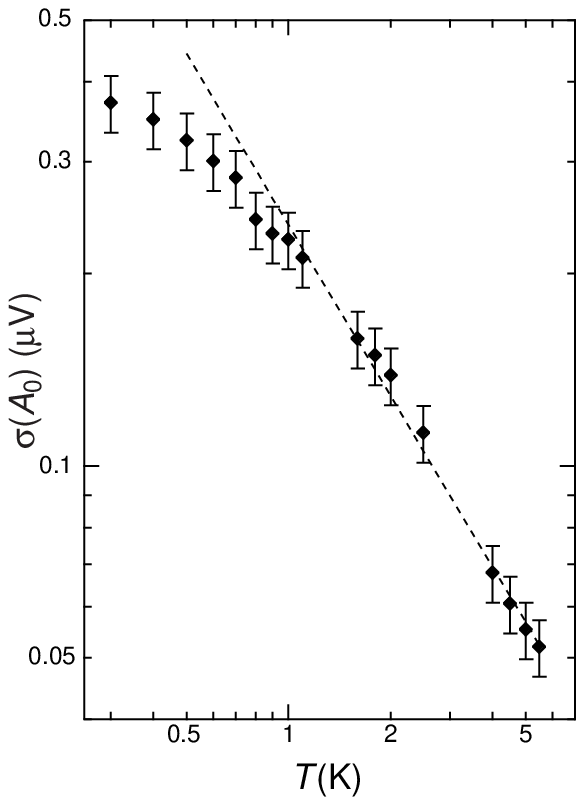}
	\caption{\figfourcaption}%
	\label{fig:fig4}
    \end{center}
\end{figure}

As pumping fluctuations extend on both sides of zero and pumping 
occurs in either direction with equal likelihood for a given $\phi$, 
\avg{\Ao} is small and the pumping amplitude is instead characterized 
by $\sAo$, the standard deviation of $\Ao$.  For example, the data in 
\figref{2}{B} yield $\avg{\Ao} = \unit[0.01]{\micro V}$ 
while the standard deviation $\sAo = \unit[0.4]{\micro V}$.  Values of 
$\sAo$ (Figs. \ref{fig:fig2}, \ref{fig:fig3}, and \ref{fig:fig4}) are 
based on 96 independent configurations over \bfield, \Vga, and \Vgb, 
(\figref{2}{B}).

The dependence of the pumping amplitude \sAo on pumping frequency is 
linear (\figref{2}).  For the above parameters, the linear 
dependence has a slope of $\unit[40]{nV/MHz}$.  As the dot has 
conductance $g \sim 2 e^{2}/h$, this voltage compensates a pumped 
current of $\unit[3]{pA/MHz}$, or roughly 20 electrons per pump cycle.  
The dependence of \sAo on the pumping strength \Aac (\figref{3}) shows 
that for weak pumping, $\Aac < \unit[80]{mV}$, \sAo is proportional to 
\Aacsq, as expected from the simple loop-area argument described 
above.  For stronger pumping \sAo increases slower than $\Aac^2$, with 
a crossover from weak to strong pumping occurring at roughly the 
characteristic gate voltage scale of fluctuations in both dot 
conductance and pumping, measured independently to be $\unit[70 \pm 
6]{mV}$.  This departure from an $\Aac^2$ dependence for strong 
pumping is expected to occur when the loop in parameter space 
describing pumping becomes sufficiently large that it encloses 
uncorrelated emissivities \cite{zhou:mesoscopic, brouwer:scattering}.  
In this case, one would expect $\sAo \propto \Aac$.  However, the 
observed dependence at strong pumping is slower then linear, and in 
fact appears consistent with $\sAo \propto \Aac^{1/2}$.  This 
unexpectedly slow dependence may result if significant heating and 
dephasing of electrons occurs due to strong pumping.  Further study is 
needed to investigate this.  Another characteristic of strong pumping is 
that $\Vdot (\phi )$ becomes nonsinusoidal, as seen in the lower inset 
of \figref{3}.  Notice that $\Vdot (\phi = \pi)$ remains close to zero 
for all pumping strengths while $\Vdot (\phi = 0)$ deviates from 0 at 
strong pumping.

The temperature dependence of \sAo for pumping strength near the 
crossover from weak to strong pumping, $\Aac = \unit[80]{mV}$ is shown 
in \figref{4}.  At high temperatures (\unit[1--5.5]{K}), \sAo is 
described by a power law, $\sAo = 0.2\, \temp^{-0.9}$ (for \sAo in 
\unit{\micro V} and \temp in \unit{K}).  This behavior presumably 
reflects the combined influence of thermal smearing, which alone is 
expected to yield $\sAo \propto \temp^{-1/2}$, and 
temperature-dependent dephasing.  A similar temperature dependence is 
found for the amplitude of conductance fluctuations in dots 
\cite{huibers:distributions}.  Below 1K, the temperature dependence 
begins to round off, perhaps indicating a saturation at lower 
temperatures.  A low-temperature saturation of pumping is expected 
when thermal smearing becomes less than lifetime broadening 
\cite{zhou:personal}, $k_BT < \left[\Gamma_{\mathrm{esc}} + 
\Gamma_{\varphi}(T)\right]$, where $\Gamma_{\mathrm{esc}} = 
N\Delta/\pi$ ($N$ is the number of modes per lead) is the broadening 
due to escape through the leads, and $\Gamma_{\varphi}(T)$ is the 
broadening due to dephasing, $\Gamma_{\varphi}(T) = 
\hbar/\tau_{\varphi}$.  Using $N\sim 2$ and known dephasing times 
$\tau_{\varphi}$ in similar dots \cite{huibers:distributions} yields 
an expected saturation at $\usim \unit[100]{mK}$, consistent with the 
rounding seen in the data.

Finally, we investigate the symmetries and statistical properties of 
adiabatic quantum pumping.  The symmetry of pumping fluctuations about 
zero magnetic field is seen in the
gray-scale plot of $\Vdot(\phi = \pi/2)$ as a function of \bfield and 
the dc voltage on one shape-distorting gate, \Vga (\figref{5}{A}).  The full symmetry 
of pumping follows from time-reversal symmetry: $\Vdot(\phi, \bfield) 
= -\Vdot(-\phi ,-\bfield)$, analogous to the Landauer-B\"{u}ttiker 
relations for conductance \cite{zhou:mesoscopic}.  The reduced 
symmetry observed in \figref{5}{A}, $\Vdot(\phi, \bfield) = 
\Vdot(\phi, -\bfield)$, results from a combination of time-reversal 
symmetry and the symmetry $\Vdot(\phi,\bfield) = 
-\Vdot(-\phi,\bfield)$ implied by the sinusoidal dependence of \Vdot 
on \ph at low pumping amplitudes.

A central paradigm in mesoscopic physics is that the statistical 
properties of a fluctuating quantity depend on symmetries of the 
system and little else.  In order to investigate how the statistics of 
pumping fluctuations depend on time-reversal symmetry, we have 
measured $\Vdot(\phi = \pi/2)$, as well as the conductance, as a 
function of magnetic field for 36 independent configurations of \Vga 
and \Vgb.  The sampling in $B$ is much finer than the characteristic 
magnetic field scales for pumping fluctuations ($\unit[3.3 \pm 
0.4]{mT}$) and conductance fluctuations ($\unit[3.9 \pm 0.4]{mT}$), 
where these values are the half-maxima of the autocorrelations of the 
fluctuations.  These values are comparable to and somewhat smaller 
than one flux quantum through the device, consistent with theory and 
previous experiments on conductance in similar dots 
\cite{marcus:quantum}.  The average pumped voltage, $\langle\Vdot(\phi 
= \pi/2)\rangle$ is close to zero and has no outstanding features 
other than its symmetry in magnetic field.  On the other hand 
$\sigma(\Vdot(\phi = \pi/2))$, shows a peak at $\bfield = 0$ of 
roughly twice its value away from zero field.  The peak width is 
comparable to the correlation field (\figref{5}{B}), suggesting that 
the peak is associated with the breaking of time-reversal symmetry.  
We conclude that pumping fluctuations are larger for the time-reversal 
symmetric case at $\bfield = 0$, similar to the situation for 
conductance fluctuations \cite{baranger:mesoscopic, 
*jalabert:universal, *chan:ballistic}.

\begin{figure}
    \begin{center}
	\includegraphics{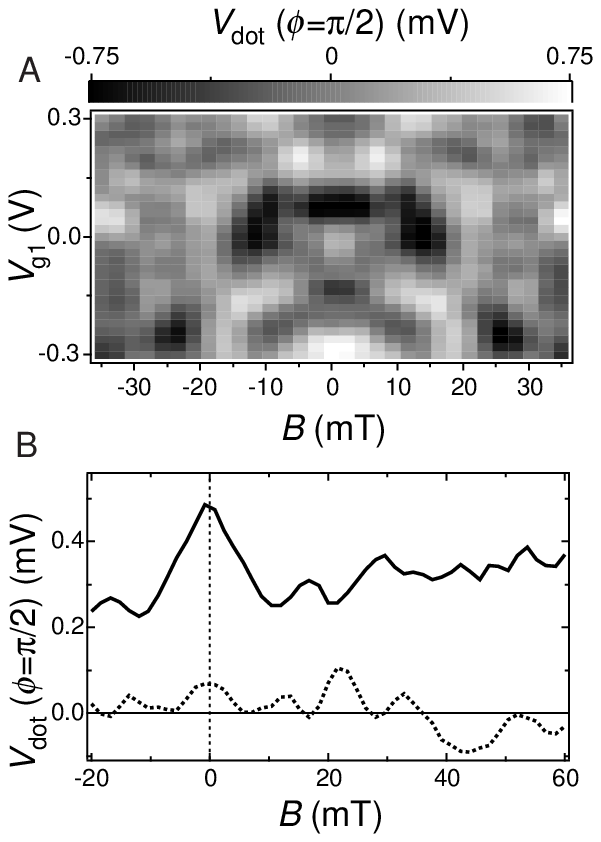}
	\caption{\figfivecaption}%
	\label{fig:fig5}
    \end{center}
\end{figure}

\iffalse{
In conclusion, we have investigated a new quantum mechanical pumping 
effect that produces dc transport in response to a cyclic deformation 
of the confining potential in phase-coherent quantum dots.  In 
agreement with theoretical predictions \cite{zhou:mesoscopic, 
brouwer:scattering}, pumping depends sinusoidally on the phase 
difference between the two shape-distorting ac voltages, and shows 
mesoscopic fluctuations about zero pumping signal (i.e.\ the direction 
of pumping varies randomly as external parameters are changed) with a 
typical amplitude that depends linearly on the pumping frequency and 
on the square of the ac driving amplitude at low amplitude.  
Unexplained aspects of this new phenomenon include the slow increase 
in pumping amplitude for strong pumping, the large exponent of the 
temperature dependence, and the ratio of typical pumping amplitude for 
broken and unbroken time reversal symmetry.  A number of issues remain 
for future work, particularly (i) the role of dephasing---in 
particular how dephasing in pumped transport compares to dephasing in 
bias-driven transport, (ii) the role of electron-electron interactions and its connection 
(through the Coulomb blockade) to the quantization of pumping 
\cite{aleiner:adiabatic}, (iii) a thorough understanding of strong 
pumping, and the appearance of dissipation due to pumping, and (iv) 
correlations between conductance and pumping, and how those 
correlations depend on the symmetry of the system 
\cite{brouwer:scattering}.\fi
\nocite{note:bfieldshift, note:thanks}


\begin{mcbibliography}{10}

\bibitem{beenakker:quantum}
   C.~W.~J. Beenakker and H.~Van~Houten, in \emph{Solid State 
   Physics}, H.~Ehrenreich and D.~Turnbull eds.  (Academic Press, San 
   Diego, 1991), vol.~44, pp.  1--228\relax \relax
\bibitem{kouwenhoven:electron}
  L.~P. Kouwenhoven, et~al., in \emph{Proceedings of the Advanced 
  Study Institute on Mesoscopic Electron Transport}, L.~P. 
  Kouwenhoven, G.~Sch\"{o}n, L.~L. Sohn eds.  (Kluwer, Dordrecht, 
  1997), Series E\relax \relax
\bibitem{thouless:quantization}
  D.~J. Thouless, \emph{Phys.  Rev.  B} \textbf{27}, 6083 (1983)\relax 
  \relax
\bibitem{spivak:mesoscopic}
  B.~Spivak, F.~Zhou, M.~T. Beal~Monod, \emph{Phys.  Rev.  B} 
  \textbf{51}, 13226 (1995)\relax \relax
\bibitem{zhou:mesoscopic}
  F.~Zhou, B.~Spivak, B.~L. Altshuler, \emph{Phys. Rev. Lett.}
  \textbf{82}, 608 (1999)\relax \relax
\bibitem{brouwer:scattering}
  P.~W. Brouwer, \emph{Phys.  Rev.  B} \textbf{58}, 10135 (1998)\relax 
  \relax
\bibitem{kouwenhoven:photon}
  L.~P. Kouwenhoven, et~al., \emph{Phys.  Rev.  B} \textbf{50}, 2019 
  (1994)\relax \relax
\bibitem{kouwenhoven:observation}
  L.~P. Kouwenhoven, et~al., \emph{Phys.  Rev.  Lett.} \textbf{73}, 
  3443 (1994)\relax \relax
\bibitem{fal:mesoscopic}
  V.~I. Fal'ko and D.~E. Khmel'nitski\u{\i}, \emph{Zh.  Eksp.  Teor.  
  Fiz.} \textbf{95}, 328 (1989) [\emph{Sov.  Phys.  JTEP} \textbf{68}, 
  186 (1989)]\relax \relax
\bibitem{liu:mesoscopic}
  J.~Liu, M.~A. Pennington, N.~Giordano, \emph{Phys.  Rev.  B} 
\textbf{45}, 1267 (1992)\relax \relax
\bibitem{kouwenhoven:quantized}
  L.~P. Kouwenhoven, A.~T. Johnson, N.~C. van~der Vaart, C.~J. P.~M. 
  Harmans, C.~T. Foxon, \emph{Phys.  Rev.  Lett.} \textbf{67}, 1626 
  (1991)\relax \relax
\bibitem{kouwenhoven:quantizedz}
  L.~P. Kouwenhoven, et~al., \emph{Z. Phys.  B} \textbf{85}, 381 
  (1991)\relax \relax
\bibitem{pothier:single}
  H.~Pothier, P.~Lafarge, C.~Urbina, D.~Esteve, M.~H. Devoret, 
  \emph{Europhys.  Lett.} \textbf{17}, 249 (1992)\relax \relax
\bibitem{keller:rare}
  M.~W. Keller, J.~M. Martinis, R.~L. Kautz, \emph{Phys.  Rev.  Lett.} 
  \textbf{80}, 4530 (1998)\relax \relax
\bibitem{buttiker:current}
  M.~B\"{u}ttiker, H.~Thomas, A.~Pr\^{e}tre, \emph{Z. Phys.  B} 
  \textbf{94}, 133 (1994)\relax \relax
\bibitem{baranger:mesoscopic}
  H.~U. Baranger and P.~A. Mello, \emph{Phys.  Rev.  Lett.} \textbf{73}, 
  142 (1994)\relax \relax
\bibitem{jalabert:universal}
  R.~A. Jalabert, J.-L. Pichard, C.~W.~J. Beenakker, \emph{Europhys.  
  Lett.} \textbf{27}, 255 (1994)\relax \relax
\bibitem{chan:ballistic}
  I.~H. Chan, R.~M. Clarke, C.~M. Marcus, K.~Campman, A.~C. Gossard, 
  \emph{Phys.  Rev.  Lett.} \textbf{74}, 3876 (1995)\relax \relax
 \bibitem{note:pumping}
   Voltages on these gates have a dc component, $V_{g}$, and an ac 
   component (at MHz frequencies) produced using two frequency-locked 
   synthesizers (HP 3325) with a computer-controlled phase difference 
   \ph between them.  To allow a sensitive lock-in measurement of the 
   pumping signal, the ac gate voltages are chopped by a low-frequency 
   (\unit[93]{Hz}) square wave, and the voltage across the dot 
   measured synchronously using a PAR 124 lock-in amplifier.  A bias 
   current can also be applied directly from the lock-in amplifier 
   allowing conductance to be measured without disturbing the 
   measurement set-up.  The current bias is then set to zero to 
   measure pumping\relax \relax
\bibitem{huibers:distributions}
  A.~G. Huibers, M.~Switkes, C.~M. Marcus, K.~Campman, A.~C. Gossard, 
  \emph{Phys.  Rev.  Lett.} \textbf{81}, 1917 (1998)\relax \relax
\bibitem{zhou:personal}
  F.~Zhou, personal communication (1998)\relax \relax
\bibitem{marcus:quantum}
  C.~M. Marcus, et~al., \emph{Chaos, Solitons, \& Fractals} 
  \textbf{8}, 1261 (1997)\relax \relax
\bibitem{note:bfieldshift}
  The line of symmetry is slightly shifted from the ``$\bfield = 0$'' 
  line determined from the magnet current due to offset magnetic 
  fields\relax \relax
\bibitem{note:thanks}
  We thank A. Auerbach, P. Brouwer, A. Morpurgo, B. Spivak, and F. 
  Zhou for useful discussions.  We acknowledge support at Stanford 
  from the ARO under contract DAAH04-95-1-0331 and the NSF-PECASE 
  under contract DMR-9629180-1, and at UCSB from the AFOSR under Grant 
  No.  F49620-94-1-0158 and by QUEST, an NSF Science and Technology 
  Center\relax \relax
\end{mcbibliography}
\end{document}